\documentstyle[12pt,fleqn]{article}
\mathchardef\SGamma="7100
\begin{document}
\title{\vskip-1.7cm \bf  The Gospel according to DeWitt revisited:
quantum effective action in braneworld models\footnote{Talk given
at Bryce DeWitt memorial session of the International Conference
on Theoretical Physics, 70 years Jubilee of the I.E.Tamm Theory
Department of the Lebedev Physics Institute, Moscow, April 11-16,
2005.}}
\date{}
\author{A.O.Barvinsky}
\maketitle \hspace{-8mm} {\em Theory Department, Lebedev Physics
Institute, Leninsky Prospect 53, Moscow 117924, Russia}

\begin{abstract}
We construct quantum effective action in spacetimes with branes
(boundaries) and establish its relation to the "cosmological wave
function" of the bulk -- the solution of the corresponding
Wheeler-DeWitt equation which can be considered as a means of the
holographic description of braneworld models. We show that for a
special type of the bulk-brane gauge fixing procedure the one-loop
part of the action decouples into the additive sum of
brane-to-brane and bulk-to-bulk effective actions, and this
decomposition proliferates in a special way in higher orders of
the Feynman diagrammatic expansion. This property is based on a
special duality relation between the Dirichlet and Neumann
boundary value problems when applied to the functional
determinants of wave operators and the field-theoretic version of
the well-known semiclassical Van Vleck-Morette determinant. It
facilitates the gauge-independent way of treating the
strong-coupling and VDVZ problems in brane induced gravity models.
Importance of this technique in various implications of braneworld
theory and infrared modifications of Einstein theory is briefly
discussed.
\end{abstract}

\section{Introduction}
This talk is dedicated to the memory of Bryce DeWitt. No need to
say that he laid foundations of quantum gravity and theory of
gauge fields. But it is worth emphasizing that he was a moving
spirit behind what later has materialized as
Bechi-Rouet-Stora-Tyutin symmetry, Batalin-Fradkin-Vilkovisky
formalism and Batalin-Vilkovisky quantization method ---
fundamental landmarks in modern high-energy physics and the whole
epoch in the life of the Theory Department whose seventy years
jubilee we are celebrating now. Moreover, for many years Bryce
DeWitt was linked with bonds of deep affection to Professor
Fradkin, Professor Markov, Igor Batalin and, especially, Gregory
Vilkovisky by sharing with them infectious enthusiasm in
challenging and attacking fundamental problems of theoretical
physics and beyond. He often visited Russia, and one can hardly
find another place here where an American professor would walk
more frequently than Bryce did in one of the streets of Moscow. It
is hard to think that he will never ever walk there again.

Huge scope of his interests and works -- the Gospel according to
DeWitt as Gregory Vilkovisky once called it \cite{gospel} --
embraced foundations of quantum mechanics, measurement theory,
quantum field theory and cosmology, and other speakers at this
session -- Richard Woodard, Bill Unruh and Slava Mukhanov -- are
going to discuss these subjects. As for me, I am going to dwell on
his pioneering ideas and methods of quantum theory of gauge
fields, and the way these methods work in the new context ---
gravitational brane models with extra dimensions. I will begin
with the peculiarities of Feynman-DeWitt-Faddeev-Popov gauge
fixing procedure \cite{DW1,DW2} in brane theory, demonstrate its
relation to the {\em cosmological wavefunction} of the
multidimensional spacetime bulk as a solution of the corresponding
Wheeler-DeWitt equation \cite{DW}, consider its semiclassical
expansion and a nontrivial interplay between the Dirichlet and
Neumann boundary value problems for propagators of the covariant
background field method in brane theory. Finally, I will briefly
discuss the significance of these methods in infrared
modifications of Einstein theory that recently got strongly
motivated by the challenge of the cosmological constant and
cosmological acceleration problems \cite{darkenergy}.

\section{Gauge fixing in braneworld models}

Motivated by this challenge, as well as by the D-brane idea in
string theory \cite{D-brane} and hirearchy problem in high-energy
physics \cite{hierarchy}, the new concept of extra dimensions
suggests the braneworld picture of our Universe as a timelike
surface embedded into the fundamental higher-dimensional
spacetime. In the simplest 5-dimensional case its canonical
description can be represented as a transformation of the familiar
picture of time evolution from an initial 3-dimensional Cauchy
surface to the picture of a timelike 4-dimensional brane with the
fifth coordinate playing the role of time. Principal difference is
that the Cauchy surface is an imaginary manifold carrying only the
initial data while the brane is an actual bearer of 4-dimensional
tension and matter spanned by its timelike trajectories which are
subject to dynamical laws.

Generically, the action in brane models contains the 5-dimensional
(bulk) and 4-dimensional (brane) parts depending on the
corresponding metrics $G_{AB}(X)$ and $g_{\alpha\beta}(x)$,
$A=0,1,2,3,5$, $\alpha=0,1,2,3$, and involves matter sources on
the brane which we do not specify here. They contain the
5-dimensional and possibly 4-dimensional Einstein terms with or
without relevant cosmological terms, and can be characterized by
different gravitational constant scales $G_4=1/M_4^2$ and
$G_5=1/M_5^3$ (like in brane induced gravity models),
    \begin{equation}
    S=S_4[\,g_{\alpha\beta}(x)\,]+S_5[\,G_{AB}(X)\,]  \label{1}
    \end{equation}

Quantum effective action results from functional integration over
the bulk and brane metrics in which, however, one should factor
out by gauge-fixing procedure spacetime diffeomorphisms,
    \begin{equation}
    e^{\,\textstyle i\SGamma}=\int
    DG_{AB}(X)\,e^{\,\textstyle iS\,[\,G_{AB}(X)\,]}
    \times(\,\mbox{gauge-fixing}\,).                  \label{2}
    \end{equation}
These diffeomorphisms, which leave the system invariant,
$G_{AB}\to G_{AB}+{\cal L}_\Xi G_{AB}$, are generated by the bulk
vector fields $\Xi^A(X)$ which do not move the brane, that is have
on the brane zero normal component
    \begin{equation}
    \Xi_\perp|=0,\,\,\,\,\Xi^\mu|
    \equiv\xi^\mu(x)\neq 0                            \label{3}
    \end{equation}
(we shall denote by the vertical bar the restriction of the
quantity to the brane). Tangential diffeomorphisms do not move the
brane and, therefore, are included in the gauge group to be
factored out.

A gauge-fixing procedure begins with adding to the action a
gauge-breaking term quadratic in gauge conditions
$F^A(G,\partial_B G)$. When gauges are relativistic, that is admit
residual transformations whose parameters form the spacetime
propagating modes, the complete gauge-fixing should include extra
gauge conditions $\chi^\mu(g,\partial_\mu g)$ imposed on the brane
to remove the degeneracy in boundary conditions for these modes.
This leads to extra gauge-breaking term as a brane-surface
integral quadratic in $\chi^\mu$, and the gauge-fixed action takes
the form
    \begin{equation}
    S\Rightarrow S_{\rm gf}=S+\frac12 F^A\,C_{AB} F^B
    +\frac12 \chi^\mu\,c_{\mu\nu}
    \chi^\nu.                               \label{4}
    \end{equation}
Here $C_{AB}$ and $c_{\mu\nu}$ are some gauge-fixing matrices and
we use condensed DeWitt notations which accumulate in the index
not only discrete labels but also spacetime coordinates, their
contraction implying also spacetime integration.\footnote{There is
a joke that in the history of the twentieth century there were
basically two achievements in theoretical physics. One is due to
Einstein who suggested to omit the summation sign and another due
to Bryce DeWitt who proposed to drop the sign of integration. And
maybe this joke is very close to a deep truth, because it is hard
to overestimate how these suggestions expedited research in
high-energy and gravitational physics.} Since $A$ is a
5-dimensional index, its condensed version $A=(A,X)$ implies
integration over the full 5D bulk, and the second term in
(\ref{4}) represents the bulk gauge-breaking term $S^{\rm gb}_5$,
while with the 4D index $\mu=(\mu,x)$ the third term is the brane
surface integral $S^{\rm gb}_4$,
    \begin{eqnarray}
    &&S^{\rm gb}_5=\frac12 F^A\,C_{AB} F^B=\int d^5X\,
    F^A(X)\,C_{AB}(X)\, F^B(X), \\
    &&S^{\rm gb}_4=\frac12 \chi^\mu\,c_{\mu\nu}
    \chi^\nu=\frac12 \int d^4x\,\chi^\mu(x)\,c_{\mu\nu}(x)\,
    \chi^\nu(x)
    \end{eqnarray}

An example of such gauge conditions is given by the linear
harmonic gauges on metric fluctuations $H_{AB}=G_{AB}-G^0_{AB}$,
$h_{\alpha\beta}=g_{\alpha\beta}-g^0_{\alpha\beta}$, on some
background (with all covariant derivatives and contractions with
respect to a background metrics)
    \begin{equation}
    F_A=\nabla^B H_{AB}-\frac12\nabla_A H,\,\,\,\,
    \chi_\mu=\nabla^\nu h_{\mu\nu}-\frac12 \nabla_\mu h.        \label{5}
    \end{equation}
The Faddeev-Popov operators for these gauges represent
correspondingly the 5D and 4D covariant d'Alembertians (modified
by Ricci curvature terms)
    \begin{equation}
    F^A\,\Rightarrow \,\textbf{Q}^A_B=
    \Box_5\delta^A_B+R^A_B,\,\,\,\,\,\,
    \chi^\mu\,\Rightarrow\,J^\mu_\nu=
    \Box_4 \delta^\mu_\nu +R^\mu_\nu        \label{6}
    \end{equation}
and thus describe the residual transformations as modes
propagating respectively in the bulk and on the brane.

More generally, if we introduce condensed notations for bulk and
brane metric fields
    \begin{equation}
    G^a=G_{AB}(X),\,\,\,g^i=g_{\alpha\beta}(x)
    \end{equation}
and denote the generators of their respective 5D and 4D
diffeomorphisms (Lie derivatives acting respectively in the bulk
and on the brane) by $R^a_A$ and $R^i_\mu$
    \begin{equation}
    {\cal L}_\Xi G^a=R^a_A\,\Xi^A,\,\,\,\,\,
    {\cal L}_\xi g^i=R^i_\mu\,\xi^\mu,
    \end{equation}
then these operators read
    \begin{eqnarray}
    &&\textbf{Q}^A_B=\frac{\delta F^A}{\delta G^a} R^a_B, \label{7a}\\
    &&J^\mu_\nu=
    \frac{\partial\chi^\mu}{\partial g^i} R^i_\mu,        \label{7}
    \end{eqnarray}
where to distinguish the 5D and 4D variational derivatives we use
correspondingly variational and partial-derivative notations.

Remarkable feature of the relativistic gauge-fixing procedure
(\ref{4}) that does not mix bulk and brane metrics is the
bulk-brane factorization of the gauge-fixing factor in (\ref{2}).
More precisely, let the brane gauges $\chi^\mu(g,\partial_\mu g)$
among the full set of gauge conditions
    \begin{eqnarray}
    &&\chi^\mu=\chi^\mu(g_{\alpha\beta},
    \partial_\mu g_{\alpha\beta}),                  \label{101} \\
    &&F^A=F^A(G_{CD},\partial_B G_{CD})             \label{100}
    \end{eqnarray}
{\em not} involve lapse-shift functions $N_A\sim G_{5A}$ and the
derivatives $\partial_5 g_{\alpha\beta}$ and let the bulk gauges
be relativistic ones from the viewpoint of the fifth-time
canonical formalism
    \begin{eqnarray}
    \det \frac{\partial F^A}{\partial\dot N^B}
    \neq 0,\,\,\,\,\dot
    N^B\equiv\partial_5 N^B.
    \end{eqnarray}
This property guarantees that the matrix-valued operator
$\textbf{Q}^A_B$, (\ref{7a}), has a non-degene\-rate term of
second order in $\partial_5$ (cf. Eq. (\ref{6})) and, thus,
describes dynamical propagation of the modes of residual
transformations (gauge ghosts) in the fifth "time".\footnote{In
its turn, this follows from the fact that the gauge
transformations of Lagrange multipliers contains the time
derivative of the gauge parameter ${\cal L}_\Xi
N^A=R^A_B\,\Xi^B\sim\partial_5\Xi^A+...$\,.} From the viewpoint of
the canonical formalism, the brane gauges (\ref{101}) play the
role of "unitary" gauge conditions with the ghost operator
(\ref{7}) ultralocal in time $x^5$, as opposed to to relativistic
gauge conditions (\ref{100}).

Then the full gauge-fixing factor in the path integral (\ref{2})
factorizes into the product of bulk and brane pieces
    \begin{eqnarray}
    (\,\mbox{gauge-fixing}\,)=\Big(\,e^{\,\textstyle \frac i2
    F^A\,C_{AB} F^B}\;
    {\rm Det}_{\bf D} \textbf{Q}^A_B \Big)\times
    \Big(\,e^{\,\frac i2\textstyle \chi^\mu\,c_{\mu\nu}
    \chi^\nu}\;\det J^\mu_\nu\ \Big),        \label{8}
    \end{eqnarray}
where we again distinguish between the notations for five
dimensional ($\rm Det$) and four dimensional ($\det$) functional
determinants of ghost operators, and the subscript {\bf D} implies
that the functional determinant of the bulk operator is calculated
subject to {\em Dirichlet boundary conditions} on the
brane.\footnote{We are not particularly concerned with the
boundary conditions for the 4D Faddeev-Popov determinant, because
they are determined by concrete (in-out or in-in) setting on the
brane and in the Euclidean case reduce to trivial Dirichlet
conditions at infinity.} This factorization allows one to rewrite
the effective action path integral (\ref{2}) in a somewhat
different form, provided the integration over brane metric is
reserved for the last, while the bulk integration (subject to
fixed induced metric on the brane) is done first according to
    \begin{equation}
    \int DG_{AB}(X)=\int
    dg_{\alpha\beta}(x)\!\!\!
    \int\limits_{\,\,\,\,\,\,
    G_{\alpha\beta}|=g_{\alpha\beta}(x)}\!
    \!\!\!\!\!\!\!\!\!DG_{AB}(X).                           \label{9}
    \end{equation}
The result looks as a purely 4-dimensional
Feynman-DeWitt-Faddeev-Popov functional integral \cite{covar}
    \begin{equation}
    e^{\textstyle i\SGamma}=\int
    dg_{\alpha\beta}(x)\,e^{\,\textstyle iS_4(\,g_{\alpha\beta}\,)}\,
    \Big(\,e^{\,\frac i2\textstyle \chi^\mu\,c_{\mu\nu}
    \chi^\nu}\,\det J^\mu_\nu\, \Big)\,
    \mbox{\boldmath$\Psi$}_{\rm Bulk}(\,g_{\alpha\beta}\,)              \label{10}
    \end{equation}
for the system with the brane action $S_4[\,g_{\alpha\beta}\,]$
but with the insertion of the, so to say, cosmological
wavefunction of the bulk spacetime
    \begin{equation}
    \mbox{\boldmath$\Psi$}_{\rm Bulk}(\,g_{\alpha\beta}\,)
    =\!\!\!\int\limits_{\,\,\,\,\,\,G_{\alpha\beta}|=g_{\alpha\beta}(x)}\!
    \!\!\!\!\!\!\!\!\!DG_{AB}(X)
    \,e^{\,\textstyle iS_5[\,G_{AB}\,]}\,
    \left(\,e^{\textstyle\,\frac i2
    F^A\,C_{AB} F^B}\,{\rm Det}_{\bf D}
    \textbf{Q}^A_B\, \right).                      \label{11}
    \end{equation}

\section{Braneworld picture vs quantum cosmology}

This is a functional of brane metric which has a number of
important properties. First it is determined only by the bulk part
of the action $S_5[\,G_{AB}\,]$. As was shown in context of the
4-dimensional quantum cosmology \cite{barv}, the path integral
(\ref{11}) is independent of the choice of bulk gauge conditions
for {\em any (off-shell)} value of its argument $g_{\alpha\beta}$
and satisfies the Wheeler-DeWitt equations \cite{DW}
--- the set of first-class Dirac constraints
$H_A(\,g_{\alpha\beta},\,p^{\alpha\beta})$ defined in fifth-time
canonical formalism and imposed in coordinate representation of
canonical commutation relations,
$p^{\alpha\beta}=\partial/i\partial g_{\alpha\beta}$,
    \begin{equation}
    H_A\Big(\,g_{\alpha\beta},
    \frac{\partial}{i\,\partial g_{\alpha\beta}}\,\Big)\,
    \mbox{\boldmath$\Psi$}_{\rm B}(\,g_{\alpha\beta}\,)=0.       \label{12}
    \end{equation}

These properties can be directly checked in the one-loop
approximation, when
    \begin{equation}
    \mbox{\boldmath$\Psi$}^{\rm 1-loop}_{\rm Bulk}(\,g\,)=
    e^{\textstyle iS_5[\,G_0(g)\,]}\,
    \frac{{\rm Det}_{\bf D} \textbf{Q}^A_B}
    {(\,{\rm Det}\, \textbf{F}_{ab}\,)^{1/2}},    \label{13}
    \end{equation}
is given by the tree-level exponential $S_5[\,G_0(g)\,]$
--- the classical bulk action on its mass shell $G_0(g)$
(the solution of classical equations in the bulk subject to
boundary data on the brane,
$G^0_{\alpha\beta}(g)|=g_{\alpha\beta}(x)$)
--- and the one-loop prefactor composed of functional determinants
of the ghost $\textbf{Q}^A_B$ and metric field $\textbf{F}_{ab}$
operators \cite{barv,Dirac} \footnote{The functional determinant
of $\textbf{F}_{ab}$ is taken subject to combined set of
Dirichlet-Neumann boundary conditions \cite{barv} which originate
from the fact that part of metric variables $G_{\mu\nu}$ are fixed
at the brane, while the lapse and shift functions $N_A\sim G_{5A}$
are integrated over in infinite limits at the brane.},
    \begin{equation}
    \textbf{F}_{ab}=\frac{\delta^2}{\delta G^a \delta
    G^b}\,\Big(\,S_5+S_5^{\rm gb}\Big).             \label{14}
    \end{equation}

In context of quantum cosmology, when the role of $x^5$ is played
by a real physical time $x^0=t$, $g_{\alpha\beta}$ goes over into
$g^i=g_{mn}({\bf x})$, $m=1,2,3$ and the gauge index $A$ shrinks
to the 4-dimensional one, $A\to\mu$, the consideration of a
one-argument wavefunction can be replaced by the two-point kernel
$\mbox{\boldmath$U$}(g,g')$,
    \begin{equation}
    \mbox{\boldmath$\Psi$}_{\rm Bulk}(\,g\,)
    \to\mbox{\boldmath$U$}(g,g'),                \label{14a}
    \end{equation}
amplitude of transition between the 3-metric configurations of two
spacelike surfaces (playing the role of two spacelike
branes).\footnote{In context of Euclidean quantum gravity the
replacement opposite to (\ref{14a}) gives rise to the well-known
no-boundary cosmological wavefunction \cite{HH} when the "initial"
3-geometry $g'$ is identified with a regular internal point (the
center) of the Euclidean spacetime ball.} As a solution of the
Wheeler-DeWitt equations with respect to $g$ (or $g'$) it has the
following semiclassical expression
    \begin{equation}
    \mbox{\boldmath$U$}_{\rm 1-loop}(g,g')=e^{\textstyle iS(g,g')}\,
    \left(\frac{\det D_{ik'}}
    {\det J^\mu_\nu\,
    \det J'^\mu_\nu\,}\,
    \right)^{1/2}                                    \label{15}
    \end{equation}
(derived in \cite{gensem} by directly solving the Wheeler-DeWitt
equations and later shown in \cite{Dirac} to be equal to the
one-loop part (\ref{13}) of the corresponding full path integral).
Here $S(g,g')$ is a Hamilton-Jacobi function (the action
calculated on classical solution interpolating between two
3-geometries $g'$ and $g$, analogous to $S_5[G_0(g_{\mu\nu})]$)
and $D_{ik'}$ is the matrix of its derivatives with respect to
initial and final metrics modified by a gauge-breaking term
    \begin{equation}
    D_{ik'}=\frac{\partial^2 S(g,g')}{\partial g^i\,\,
    \partial g^{k'}}
    +\frac{\partial\chi^\mu(g)}{\partial\, g^i}\,c_{\mu\nu}
    \frac{\partial\chi^\nu(g')}{\partial\, g^{k'}}.      \label{16}
    \end{equation}

The preexponential factor here can have the name of Morette-DeWitt
determinant, because it makes the synthesis of the well-known Van
Vleck-Morette formula for the semiclassical kernel of the
Schroedinger evolution in a non-gauge theory \cite{Morette}
    \begin{equation}
    \left[\,{\rm Det}\,\frac{\delta^2 S[\,g(t)\,]}{\delta g(t)\,
    \delta g(t')}\,\right]^{-1/2}={\rm const}\,
    \left[\,\det\,\frac{\partial^2 S(g,g')}{\partial g\;
    \partial g'}\,\right]^{1/2}                    \label{17}
    \end{equation}
with the Feynman-DeWitt-Faddeev-Popov gauge-fixing procedure in a
gauge one.  From the viewpoint of the latter the Hessian matrix of
the Hamilton-Jacobi function $S(g,g')$ is degenerate with the
gauge generators as left and right zero vectors
    \begin{equation}
    R^i_\mu\,\frac{\partial^2 S}{\partial g^i\,
    \partial g^{k'}}=0,\,\,\,\,
    \frac{\partial^2 S}{\partial g^i\,
    \partial g^{k'}}\,R^{k'}_\nu=0.                \label{18}
    \end{equation}
The gauge-breaking term in (\ref{16}), built of gauge conditions
associated with these invariances, makes this matrix invertible,
but requires introduction of the ghost factors in (\ref{15}) with
    \begin{equation}
    J^\mu_\nu=
    \frac{\partial\chi^\mu(g)}{\partial\, g^i}\,R^i_\nu,\,\,\,\,
    {J'}^\mu_\nu=
    \frac{\partial\chi^\nu(g')}{\partial\, g^{k'}}
    \,R^{k'}_\nu                                         \label{19}
    \end{equation}
in order to preserve Ward identities for the one-loop solution (in
the form of its independence of the choice of gauges $\chi^\mu$).
Thus, the equality of one-loop (\ref{13}) and semiclassical
(\ref{15}) representations is a direct gauge-theory analogue of
the non-gauge Van Vleck-Morette formula (\ref{17}).\footnote{One
might say that the Morette-DeWitt formula (\ref{15}) unifies the
research efforts of Bryce DeWitt and his wife Cecile Morette who,
as she joked when being awarded the Marcel Grossmann Prize on
General Relativity, never collaborated on any research project but
their four daughters.}

Semiclassical solutions of the Wheeler-DeWitt equations
$\mbox{\boldmath$\Psi$}_{1,2}(\,g\,)$
    \begin{equation}
    H_\mu\Big(\,g,
    \frac{\partial}{i\,\partial g}\,\Big)\,
    \mbox{\boldmath$\Psi$}_{1,2}(\,g\,)=0       \label{20}
    \end{equation}
propagated by the two-point kernel (\ref{15}) conserve the innner
product involving a nontrivial integration measure which in the
semiclassical approximation reads as \cite{gensem}
    \begin{equation}
    \langle{\bf\Psi_1},\,{\bf\Psi_2}\rangle=\int dg\,
    {\bf\Psi_1}^*(g)\,\Big(\delta(\chi(g))\,
    \det J^\mu_\nu\Big)\,{\bf\Psi_2}.      \label{21}
    \end{equation}
This measure is located on the surface of {\em unitary} gauge
conditions via the delta-function,
$\delta(\chi(g))=\prod_\mu\delta(\chi^\mu(g))$, or can be
identically transformed to the Gaussian distribution
    \begin{equation}
    \delta(\chi(g))\to (\det c_{\mu\nu})^{1/2}\;
    e^{\,\frac i2\textstyle
    \chi^\mu\,c_{\mu\nu}\chi^\nu}              \label{22}
    \end{equation}
smeared near $\chi^\mu=0$.\footnote{Generally the transformation
(\ref{22}) is identical only in the singular limit
$c_{\mu\nu}\to\infty$, but in virtue of the gauge-independence
properties of the solutions of Wheeler-DeWitt equations (\ref{20})
this replacement is identical for {\em any} invertible
$c_{\mu\nu}$. Everywhere above we neglected the determinants $\det
c_{\mu\nu}$ and ${\rm Det}\,C_{AB}$ -- the contributions of the
so-called Nielsen-Kallosh ghosts, which in field theories are
trivial in the class of ultralocal gauge-fixing matrices.}

Therefore, returning to the braneworld context and comparing
(\ref{21}) with (\ref{10}), one can say the brane effective action
(\ref{10}) can be viewed as matrix element between the wave
function of the brane ${\bf\Psi}_{\rm brane}(g)=\exp[-iS_4(g)]$
and that of the bulk (\ref{11})
    \begin{equation}
    e^{\,\textstyle i\SGamma}=
    \langle{\bf\Psi}_{\rm brane},\,
    {\bf\Psi}_{\rm Bulk}\rangle .             \label{23}
    \end{equation}
It is important to notice, though, that while in 4D quantum
cosmology all four diffeomorphisms are factored out by gauge
conditions in the measure, in 5D braneworld model only
4-dimensional diffeomorphisms are gauged away in (\ref{10}). This
corresponds to the fact that the motion of the brane is a
dynamical process, in contrast to quantum cosmology where local
deformations of the spacelike surface is a gauge transformation.

The practical meaning of the representation (\ref{23}) is still to
be comprehended, but apparently it can be utilized in
nonperturbative methods based on solving the Wheeler-DeWitt
equation for ${\bf \Psi}_{\rm Bulk}(g_{\alpha\beta})$. At the
moment, I know only of the tree-level application of this equation
by Verlindes \cite{Verlinde} who analyzed in the language of the
Einstein-Hamilton-Jacobi function the AdS-flow in the holographic
description of the Randall-Sundrum model. Also it can arise in
Gutperle-Strominger context of S-branes \cite{GutStrom} --- the
situation most closely related to chronologically ordered
mediation by quantum bulk of the interaction between S-brane
states (spacelike branes) in the low-energy limit of string field
theory.\footnote{One should notice the unusual sign of phase in
the definition of ${\bf\Psi}_{\rm brane}(g)$ above. Point is that
in contrast to conserved inner product in quantum cosmology
(\ref{21}) no unitarity holds in the fictitious propagation in
spacelike $x^5$-time, so that the inner-product representation
(\ref{23}) looks currently more as a calculational trick, rather
than an ample bulk-to-brane transition amplitude. In Euclidean
braneworld theory no such problem arises, but the formal link with
unitarity and probability conservation disappears at fundamental
level. However, in S-brane context unitary transition amplitudes
between various spacelike branes (mediated by the bulk) should
make sense. Future string field theory will apparently upgrade its
D and S-brane descriptions to a unified unitary framework in
Lorentzian spacetime. Related discussion of this point can be
found in \cite{Wickrotation}.} Postponing, however, these
nonperturbative aspirations till, hopefully, not so distant
future, let us focus on the semiclassical expansion of brane
effective action.

\section{Semiclassical expansion and duality of boundary value
problems}

Brane action was intensively studied at the tree level in various
models like Randall-Sundrum \cite{RS}, Gregory-Rubakov-Sibiryakov
\cite{GRS}, brane induced gravity models of the
Dvali-Gaba\-dadze-Porrati type \cite{DGP} for the purpose of
finding a consistent infrared modification of Einstein theory,
that could account for the phenomenon of the recently observed
cosmological acceleration \cite{darkenergy}. These models were
shown to suffer from a number of problems like van
Dam-Veltman-Zakharov discontinuity violating the correspondence
principle with the Einstein theory \cite{VDVZ}, low
strong-coupling scale precluding from consistent weak-field
perturbation theory \cite{strong,covar}, presence of ghosts and
tachyons, etc., and their eradication at the tree level did not
guarantee the consistency of the theory at the quantum level. On
the other hand, naive Feynman loop calculations lead to
uncontrollable gauge dependence of quantum effects, the lack of
their manifest covariance and other difficulties \cite{covar}.

The way to their resolution was, however, very well presented in
lessons given to us many years ago by Bryce DeWitt. Not only did
he suggest the relativistic gauge-fixing procedure of the above
type, but also invented the method of background covariant gauges
which make the background (mean) field method manifestly covariant
\cite{DeWitt}. Interestingly, applying the combination of these
methods within a special gauge fixing procedure discussed above
leads to the following one-loop approximation for brane effective
action
    \begin{equation}
    e^{\,\textstyle i\SGamma_{\rm 1-loop}(g_{\alpha\beta})}=
    e^{\,\textstyle iS[\,G_0(g_{\alpha\beta})\,]}\,
    \frac{{\rm Det}
    \textbf{Q}}{\big({\rm Det}\,{\bf F}_{\rm BB}\big)^{1/2}}\,
    \frac{\det J}
    {\big(\det{F}_{\rm bb}\big)^{1/2}},             \label{24}
    \end{equation}
where ${\bf F}_{\rm BB}$ and $F_{\rm bb}$ schematically denote the
bulk-to-bulk and brane-to-brane inverse propagators of the theory
given respectively by (\ref{14}) and
    \begin{equation}
    ({F}_{\rm bb})_{ik}=\frac{\delta^2}{\delta g^i \delta
    g^k}\,\Big(\,S_4^{\,\,\rm gf}(\,g\,)
    +S_5^{\,\,\rm gf}[\,G_0(g)\,]\Big)             \label{25}
    \end{equation}

Remarkable property of this answer is a factorization of the
prefactor into bulk-to-bulk and brane-to-brane parts which takes
place not only in the ghost sector, like in (\ref{8}), but in the
metric field sector as well. As a result the one-loop action
becomes an additive sum of bulk and brane pieces. They both
contain their respective 5D and 4D metric field and ghost
contributions with 5D relativistic Faddeev-Popov operator
$\textbf{Q}$ vs its "unitary" counterpart on the brane $J$.
Moreover, they both are separately gauge-independent. The gauge
independence of the bulk part is universal, because in the
braneworld setting there are no sources in the bulk, and the bulk
part is always on shell. The brane part is gauge-independent in
the usual sense, when the background (mean) field on the brane
$g_{\alpha\beta}(x)$ is on shell --- that is for scattering
problems on the brane.\footnote{The mean (background) field
dependence of $\SGamma(g_{\alpha\beta})$ in (\ref{24}) results
from introducing the sources to quantum fields on the right-hand
side of Eqs. (\ref{2}) and (\ref{10}) and making the Legendre
transform from the resulting generating functional of connected
Green's functions to the effective action of the mean field
$g_{\alpha\beta}(x)$ --- the generating functional of one-particle
irreducible diagrams. For brevity we did not do that in equations
throughout previous sections.} The gauge-independent off-shell
extension of this part of the action can be done along the lines
of the so-called unique effective action by G.Vilkovisky
\cite{unique,DWunique}.

The factorization of metric-field contributions in (\ref{24}) is a
corollary of an interesting duality between the Dirichlet and
Neumann boundary value problems \cite{duality} for the bulk
operator ${\bf F}$. The latter problem naturally arises in the
original definition of the brane effective action (\ref{2}). The
integration in (\ref{2}) runs also over the boundary metric, which
technically results in the Israel junction conditions on the brane
\cite{Israel}
--- the generalized Neumann or Robin type boundary condition for
quantum perturbations, their Neumann Green's function ${\bf
G}_{\bf N}$ and the corresponding one-loop functional determinant
${\rm Det}_{\bf N}{\bf F}$.

It turns out that the Dirichle-Neumann duality relates this
Green's function to the brane-to-brane operator $\textbf{F}_{\rm
bb}$ and allows one to reduce the functional determinants subject
to Neumann conditions to that of the Dirichlet ones
    \begin{eqnarray}
    &&{F}_{\rm bb}=
    \Big(\,{\bf G}_{\bf N}||\,\Big)^{-1},  \label{26}\\
    &&{\rm Det}_{\bf N}{\bf F}=
    {\rm Det}_{\bf D}{\bf F} \times \det {F}_{\rm bb}. \label{27}
    \end{eqnarray}
The double vertical bar here implies the restriction of the both
arguments of ${\bf G}_{\bf N}(X,X')$ to the boundary, after which
it becomes a kernel of the nonlocal operation on the surface,
${\bf G}_{\bf N}(X,X')||={F}_{\rm bb}^{-1}(x,x')$, which is just
the brane-to-brane propagator \cite{duality}. Again, this relation
can be regarded as an extension of the Van Vleck-Morette formula
(\ref{17}), and its application leads to the factorization
(\ref{24}) in question.

Such factorization does not literally extend beyond one-loop
order, but its elements proliferate to multi-loop Feynman diagrams
as well. They turn out to be composed of purely bulk loops, built
of Dirichlet-type bulk propagators, brane-to-brane loops and
special propagators joining them -- the structure subject to
manageable analysis in which manifest covariance and
gauge-independence are strictly kept under control. In fact, the
presence of the brane contents (characterized by the brane part
$S_4[\,g_{\alpha\beta}(x)\,]$ of the full action (\ref{1})) enters
the full quantum effective action as brane-dependent insertions in
the purely bulk Feynman diagrams. The latter contain as metric and
ghost field propagators only the Green's functions subject to
Dirichlet boundary conditions\footnote{In essence, the boundary
conditions in metric sector are slightly more complicated, cf.
footnote 4, but they anyway do not depend on the brane part of the
full classical action.} independent of
$S_4[\,g_{\alpha\beta}(x)\,]$. This is very important in models
with two different scales, like brane induced gravity models with
brane and bulk Planckian scales $M_4^2\gg M_5^2$ \cite{DGP},
because the action decouples the purely bulk part which does not
feel additional scale $M_4$ associated with the brane. This
property can be very helpful in infrared modifications of Einstein
gravity theory and other models encumbered with the
strong-coupling problem \cite{covar}.

On the other hand, duality relations (\ref{26})-(\ref{27}) allow
one to systematically reduce complicated boundary conditions of
the generalized Robin type to much simpler Dirichlet ones and
apply the well-known Schwinger-DeWitt technique of the curvature
expansion in the background field formalism of the brane effective
action --- the programme to be realized in a forthcoming paper
\cite{work}. In particular, these relations are expected to
simplify the technique for boundary surface contributions to
Schwinger-DeWitt coefficients, which can be very complicated for
Robin-type boundary conditions \cite{Vasilevich} and, especially,
for the so-called oblique boundary conditions involving the
derivatives of fields tangential to the boundary
\cite{OsbornMcAvity} and often resulting in the strong ellipticity
problem \cite{ellipticity,ellipticity1}.

To summarize, the methods of quantum brane effective action,
discussed above, are expected to be very productive in infrared
modifications of Einstein theory, stringy D-branes, (boundary)
string field theory and so on. There are many, many more
revelations from the Gospel according to Bryce DeWitt in our
coming scientific endeavors!

\section*{Acknowledgements}

This work was supported by the Russian Foundation for Basic
Research under the grant No 05-01-00996 and the LSS grant No
1578.2003.2.


\begin{thebibliography}{99}
\bibitem{gospel}G.A.Vilkovisky, The Gospel according to DeWitt, in
{\em Quantum Theory of Gravity}, ed. S.M.Christensen (Bristol,
Adam Hilger,1984).

\bibitem{DW1}B.S.DeWitt, {\em Dynamical Theory of Groups and
Fields} (Gordon and Breach, New York, 1965).

\bibitem{DW2}B.S.DeWitt, Phys.Rev. {\bf 162} (1967) 1195.

\bibitem{DW}B.S.DeWitt, Phys.Rev. {\bf 160} (1967) 1113.

\bibitem{darkenergy}A.G.Riess et al., Astron. J. {\bf 116} (1998) 109;
S.Perlmutter et al., Astrophys. J. {\bf 517} (1999) 565.

\bibitem{D-brane}J.Polchinsky, Phys. Rev. Lett.  {\bf 75} (1995)
4724, hep-th/9510017; TASI lectures on D-branes, hep-th/9611050.

\bibitem{hierarchy}L.Randall and R.Sundrum, Phys. Rev. Lett. {\bf
83} (1999) 3370, hep-ph/9905221.

\bibitem{covar}A.O.Barvinsky, Phys. Rev. {\bf D71} (2005) 084007,
hep-th/0501093.

\bibitem{barv}A.O.Barvinsky, Phys. Lett {\bf 195 B} (1987) 344.

\bibitem{Dirac}A.O.Barvinsky, Nucl. Phys. {\bf B 520} (1998) 533.

\bibitem{HH}J.B.Hartle and S.W.Hawking, Phys.Rev. {\bf D28}
(1983) 2960.

\bibitem{gensem}A.O.Barvinsky, Phys. Lett {\bf 241 B} (1990) 201.

\bibitem{Morette}C.DeWitt-Morette, Ann.Phys. {\bf 97} (1976) 367.

\bibitem{Verlinde}E.Verlinde and H.Verlinde,
%``RG-flow, gravity and the cosmological constant,''
JHEP {\bf 0005} (2000) 034, hep-th/9912018.

\bibitem{GutStrom}M.Gutperle and A.Strominger, JHEP {\bf 0204} (2002)
018, hep-th/0202210.

\bibitem{Wickrotation}A.Sen, JHEP {\bf 0204} (2002) 048, hep-th/0203211;
M.R.Gaberdiel, M.Gutperle, JHEP {\bf 0502} (2005) 051,
hep-th/0410098

\bibitem{RS}L.Randall, S.Sundrum, Phys. Rev. Lett.
{\bf 83} (1999) 4690, hep-th/9906064.

\bibitem{GRS}R.Gregory, V.A.Rubakov and S.M.Sibiryakov, Phys. Rev.
Lett. {\bf 84} (2000) 5928, hep-th/0002072.

\bibitem{DGP}G.Dvali, G.Gabadadze and M.Porrati, Phys. Rev. Lett.
{\bf B485} (2000) 208, hep-th/0005016.

\bibitem{VDVZ}H. van Damm and M.J.Veltman, Nucl. Phys. {\bf B22}
(1970) 397; V.I.Zakharov, JETP Lett. {\bf 12} (1970) 312.

\bibitem{strong}M.A.Luty, M.Porrati and R.Ratazzi, JHEP {\bf 0309}
(2003) 029, hep-th/0303116.

\bibitem{DeWitt}B.S.DeWitt, in {\em Quantum Gravity 2}, eds. C.J.Isham,
R.Penrose and D.Sciama (Oxford University Press, Oxford, 1981).

\bibitem{unique}G.A.Vilkovisky, Nucl. Phys. {\bf B234} (1984) 125.

\bibitem{DWunique}B.S.DeWitt, The Effective Action, in {\em Quantum
Field Theory and Quantum Statistics}, eds. I.A.Batalin, C.J.Isham
and G.A.Vilkovisky (Bristol, Adam Hilger,1987).

\bibitem{Israel}W.Israel, Nuovo Cim. {\bf B 44S10} (1966) 1.

\bibitem{duality}A.O.Barvinsky and D.V.Nesterov, Nucl. Phys.
{\bf B654} (2003) 225, hep-th/0210005.

\bibitem{work}A.O.Barvinsky, work in progress.

\bibitem{Vasilevich}D.V.Vassilevich, Phys. Rept. {\bf 388} (2003)
279, hep-th/0306138.

\bibitem{OsbornMcAvity}D.M.McAvity and H.Osborn,
Class. Quant. Grav. {\bf 8} (1991) 1445.

\bibitem{ellipticity}P.Gilkey, {\em Invariance Theory, the Heat
Equation, and the Atia-Singer Index Theorem} (CRC Press, Boca
Raton, FL, 1995).

\bibitem{ellipticity1}I.G.Avramidy and G.Esposito,
Class. Quant. Grav. {\bf 15} (1998) 1141, hep-th/9708163;
G.Esposito, G.Fucci, A.Yu.Kamenshchik, K.Kirsten, Class. Quant.
Grav. {\bf 22} (2005) 957, hep-th/0412269.





\end{thebibliography}
\end{document}